\documentclass[prb,floatfix,twocolumn,showpacs,amsmath,amssymb]{revtex4}
\usepackage{graphicx}
\usepackage{dcolumn}
\usepackage{bm}
\begin{document}

\title{Phase diagram of hard-core bosons on a frustrated zig-zag ladder}
\author{Davide Rossini}
\affiliation{NEST, Scuola Normale Superiore \& Istituto di Nanoscienze - CNR, Piazza dei Cavalieri 7, I-56126 Pisa, Italy}
\author{Valeria Lante}
\affiliation{Dipartimento di Fisica e Matematica, Universit\`a dell'Insubria, Via Valleggio 11, I-22100 Como, Italy}
\author{Alberto Parola}
\affiliation{Dipartimento di Fisica e Matematica, Universit\`a dell'Insubria, Via Valleggio 11, I-22100 Como, Italy}
\author{Federico Becca}
\affiliation{Democritos Simulation Center CNR-IOM Istituto Officina dei Materiali, Via Bonomea 265, I-34136, Trieste, Italy}

\date{\today} 

\begin{abstract}
We study hard-core bosons with unfrustrated nearest-neighbor hopping $t$ and 
repulsive interaction $V$ on a zig-zag ladder. As a function of the boson 
density $\rho$ and $V/t$, the ground state displays different quantum phases.
A standard one-component Tomonaga-Luttinger liquid is stable for $\rho<1/3$ 
(and $\rho>2/3$) at any value of $V/t$. At commensurate densities $\rho=1/3$, 
$1/2$, and $2/3$ insulating (crystalline) phases are stabilized for a 
sufficiently large interaction $V$. For intermediate densities $1/3<\rho<2/3$ 
and large $V/t$, the ground state shows a clear evidence of a bound state of 
two bosons, implying gapped single-particle excitations but gapless excitations
of boson pairs. Here, the low-energy properties may be described by a 
two-component Tomonaga-Luttinger liquid with a finite gap in the antisymmetric
sector. Finally, for the same range of boson densities and weak interactions, 
the system is again a one-component Tomonaga-Luttinger liquid with no evidence
of any breaking of discrete symmetries, in contrast to the frustrated case,
where a ${\cal Z}_2$ symmetry breaking has been predicted.
\end{abstract}

\pacs{}

\maketitle

\section{Introduction}

Quantum many-body systems can give rise to remarkable collective states of 
matter that have no counterpart in their classical analogs. Archetypal 
examples include superfluids, superconductors, and insulating quantum liquids 
that play a major role in modern Condensed Matter and Materials Science. 
However, collective quantum phenomena are also ubiquitous in other contexts, 
like Nuclear Physics, Quantum Chemistry, and Atomic Physics. In particular, 
ultra-cold atoms loaded into optical lattices provide a unique possibility 
for engineering quantum systems with a very high degree of tunability and 
control of the experimental parameters.~\cite{blochrev} They allow the 
realization of ``quantum simulators'' for ideal Condensed Matter models, which
may provide answers to fundamental questions.~\cite{lewenstein}
The first striking demonstration in this direction has been the observation 
of the superfluid to Mott insulator transition for bosons with short-range 
interactions;~\cite{bloch} more recently, a fermionic Mott insulator has been
also observed.~\cite{esslinger} Theoretical progresses and experimental 
achievements steadily open new research directions. Highly non-trivial 
phenomena and very rich phase diagrams are now conceivable by considering 
further ingredients, like long-range interactions, spin or multi-species 
models, frustration, and disorder.

Cold gases of bosonic particles trapped in optical lattices may be very well 
described by simple Bose-Hubbard models,~\cite{fisher} which contain hopping 
and short-range interaction terms.~\cite{jaksch,demler} Models of 
strongly-interacting bosons in one-dimensional (1D) or quasi-1D lattices 
constitute important examples where unconventional phases can be stabilized at 
low temperature. These systems do not represent a purely abstract problem, 
since it is now possible to confine the atomic species in almost decoupled 1D 
tubes, with~\cite{paredes} or without~\cite{kinoshita} an optical lattice. 
In realistic experimental setups, beyond pure 1D systems, the case of a two-leg
ladder is very easy to obtain. Indeed, one can realize a double-well potential
along a direction (say, $y$) like in Ref.~\onlinecite{oberthaler}, and a 
potential creating a cigar geometry in the $x$-axis. Then, by superimposing a 
further periodic potential along $x$, one realizes a two-leg Hubbard model. 
By playing with the distance between tubes and the height of the barrier 
between the two legs, one could tune the hopping rate between the legs. 
Likewise, the intra-chain hopping rate can be tuned by appropriately setting 
the strength of the periodic potential along the $x$ direction.

Recent experimental results have driven a new impetus to understand their 
relevant low-energy properties. In particular, the study of quasi-1D systems, 
e.g., ladders, may be very important, in order to elucidate the nature of 
exotic quantum phases that escape the standard Tomonaga-Luttinger 
theory.~\cite{giamarchi} From a theoretical point of view, ladders are 
quasi-one-dimensional systems. For this kind of anisotropic systems, 
very efficient and accurate numerical methods have been also developed 
in the last 20 years, like exact diagonalizations by the Lanczos 
technique~\cite{lanczos}, or the density-matrix renormalization group (DMRG) 
method.~\cite{white} Within these approaches, it is possible to have 
numerically exact results on fairly large clusters, so to have insights into
the physical properties at the thermodynamic limit.
Moreover, in 1D systems a clear theoretical framework is provided by the 
bosonization method,~\cite{giamarchi} which is helpful for classifying 
the possible phases.

\begin{figure}
\includegraphics[width=0.90\columnwidth]{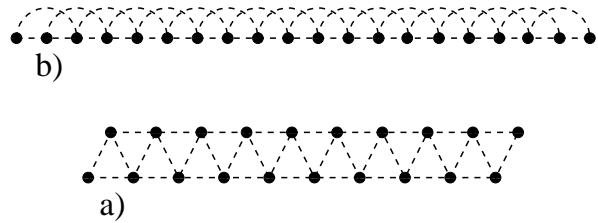}
\caption{\label{fig:ladders}
The two-leg ladder (a) is topologically equivalent to a one-dimensional chain 
with nearest and next-nearest connections (b).} 
\end{figure}

In the following, we will consider a zig-zag ladder with two legs, which is 
topologically equivalent to a 1D lattice with equal nearest and next-nearest 
neighbor hopping and interaction, see Fig.~\ref{fig:ladders}. We will study
the case of hard-core bosons that interact through a nearest-neighbor 
potential:
\begin{equation}\label{eq:hamilt}
{\cal H} = -t \sum_{\langle i,j \rangle} ( b^\dagger_i b_j + h.c. )
+V \sum_{\langle i,j \rangle} n_i n_j,
\end{equation} 
where $\langle i,j \rangle$ indicate nearest-neighbor sites in the zig-zag
geometry of Fig.~\ref{fig:ladders}, $b^\dagger_i$ ($b_i$) creates (destroys) a 
boson on the site $i$ and $n_i=b^\dagger_i b_i$ is the boson density. 
The hard-core constraint is imposed via the additional requirement $n_i =0,1$ 
on each site. The number of sites and bosons will be denoted by $L$ and $M$, 
respectively. The boson density will be denoted by $\rho=M/L$. 
In the following, we will focus on the case of unfrustrated hopping, i.e.,
$t>0$.

We mention that the Hamiltonian~(\ref{eq:hamilt}) maps onto a system of $S=1/2$ 
spins ($S^z_i=n_i-1/2$) with antiferromagnetic coupling $J^z=V$ between 
the $z$ components of spins and super-exchange $J^{xy}=-2t$ between their 
$x$ and $y$ components. Previous works have been focused on the case with 
negative hoppings, i.e., $t<0$, that corresponds to an antiferromagnetic 
Heisenberg model. In particular, the one-dimensional chain with SU(2) symmetry
(i.e., $V=2|t|$) but different nearest- ($J_1$) and next-nearest-neighbor 
($J_2$) interactions has been widely discussed, also in presence of a finite 
magnetic field.~\cite{okunishi,vekua,hikihara}
On the other hand, here we are interested in the case with positive hopping 
parameters and $V>2t$, in order to describe strongly interacting bosons in
low-dimensional systems that are relevant for atomic gases trapped in optical 
lattices. However, we will show that some features of the phase diagram
do not depend upon the sign of $t$ and can be understood on the basis of the 
strong-coupling (classical) limit $V/t \to \infty$.

For future reference, it is useful to point out that in lattices where each
site has the same coordination number $z$ (equal to the number of neighbors),
the microscopic model is invariant under ``particle-hole symmetry'', defined 
as the spin rotation of an angle $\pi$ around the $x$-axis. This particle-hole 
transformation, which is indeed a canonical transformation due to the hard-core
constraint, allows one to relate the full energy spectrum at different 
densities:
\begin{equation}\label{eq:ph}
E(L-M) = E(M) + \frac{Vz}{2}\, (L-2M)
\end{equation}
giving rise to a phase diagram symmetrical across the line $\rho=1/2$.
Open-boundary conditions violate this symmetry leading to specific finite-size
effects, since a few sites have a smaller coordination number. In the
following, open-boundary conditions will be used in DMRG, while periodic
boundary conditions will be adopted in Lanczos calculations.

\begin{figure}
\includegraphics[width=0.90\columnwidth]{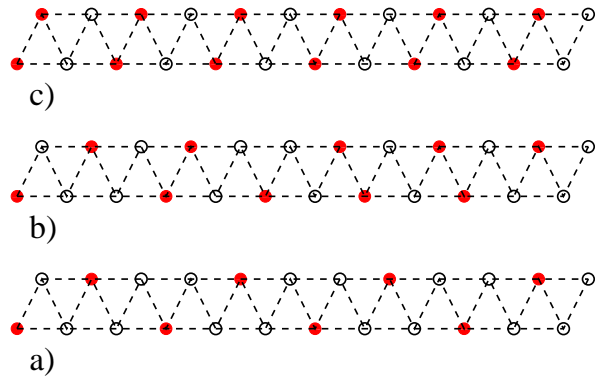}
\caption{\label{fig:twolegV}
(Color online) Examples of classical ground states for different boson 
densities. Full (empty) circles indicate particles (empty sites). 
(a) $\rho=1/3$, where the ground state is three-fold degenerated; 
(b) $1/3<\rho<1/2$, where the ground state has a finite entropy; 
(c) $\rho=1/2$, where the ground state is four-fold degenerated and corresponds
to pairs of nearest-neighbor particles.} 
\end{figure}

The paper is organized as follows: in Sec.~\ref{sec:preliminary}, we illustrate
some useful preliminary consideration related to the classical limit of $t=0$;
in Sec.~\ref{sec:results}, we present our numerical results; finally, in
Sec.~\ref{sec:conclusions}, we draw our conclusions.

\section{Preliminary considerations}\label{sec:preliminary}

Before presenting our numerical results, we would like to discuss in some 
detail the classical limit of $t=0$, which is relevant for the physical
properties in the strong-coupling regime, i.e., $V/t \gg 1$. Let us take for 
convenience $L=3 \times n$ with $n$ multiple of $4$ and quantize the $q$ 
vectors as $q=\frac{2\pi}{L} k$, with $k=0,\dots, L-1$. Whenever the number of
bosons $M<n$, i.e., $\rho=M/L<1/3$, the classical ground state has 
zero energy and finite entropy. A finite value of $t/V$ will drive the system 
towards a Tomonaga-Luttinger liquid with long-range charge-density-waves or 
superfluid correlations. Exactly at $M=n$ (namely $\rho=1/3$) a commensurate 
density wave of longitudinal wave-vector $q=2\pi/3$ sets in, leading to a 
three-fold degenerate solid-like ground state, as shown in 
Fig.~\ref{fig:twolegV}(a). The energy gap is proportional to $V$ and, 
therefore, this insulating state is expected to remain stable also in presence
of a small hopping parameter (see below, our numerical results). By further 
adding bosons, we find that for any density of the form $M=n+2 \delta$, with 
$\delta$ equal to an integer smaller than $n/4$ (so that $1/3<\rho<1/2$), the 
ground state is highly degenerate and still gapped, with energy $E=3V \delta$ 
and single particle gap $E(M+1)-E(M)=2V$. All possible ground states may be 
obtained by viewing the state, in the one dimensional topology, as a 
mixture of single particles and nearest-neighbor pairs, and placing these 
objects on the lattice in such a way that the nearest-neighbor sites of each 
object are empty, see Fig.~\ref{fig:twolegV}(b). 
Given the huge degeneracy of all these classical states, a finite value of 
$t$ has a dramatic effect, as it will be shown in the next section. Note
that for these densities the binding energy $\Delta=E(M)+E(M+2)-2\,E(M+1)$ is 
negative (i.e., $\Delta=-V$) implying the formation of boson pairs in the 
model. Finally, for $\rho=1/2$, only nearest-neighbor pairs are present and 
the ground state is a four-fold degenerate gapped ``molecular solid'' 
characterized by a crystal ordering with longitudinal wave-vector $q=\pi/2$,
see Fig.~\ref{fig:twolegV}(c).

These results for the two-leg ladder in the classical limit can be conveniently
summarized by considering the boson density $\rho$ as a function of the 
chemical potential $\mu$: at $\mu=0$, the density jumps from $\rho=0$ to 
$\rho=1/3$, where it displays a plateau up to $\mu=3V/2$. Then, the density 
jumps again to $\rho=1/2$ and then remains constant up to $\mu=5V/2$. 
A density discontinuity from $\rho=1/2$ to $\rho=2/3$ is followed by a plateau
to $\mu=4V$ where $\rho$ jumps again to its limiting value $\rho=1$. 
Therefore, in the classical limit, the model displays three distinct 
gapped (solid) phases of density $\rho=1/3$, $1/2$, and $2/3$, besides the 
trivial ``empty'' ($\rho=0$) and ``full'' ($\rho=1$) states.  
At intermediate densities phase coexistence between neighboring phases sets in. 
As we will see in the next paragraph, a finite value of $t$ removes this 
coexistence by favoring stable phases. In particular, for $1/3<\rho<1/2$, 
long-range charge-density-wave correlations develop at an incommensurate 
wave-vector $q$, which smoothly interpolates between the two limiting values 
($q=2\pi/3$ for $\rho=1/3$, and $q=\pi/2$ for $\rho=1/2$) characterizing the 
two solids at coexistence for $t=0$. 

\begin{figure}
\includegraphics[width=0.90\columnwidth]{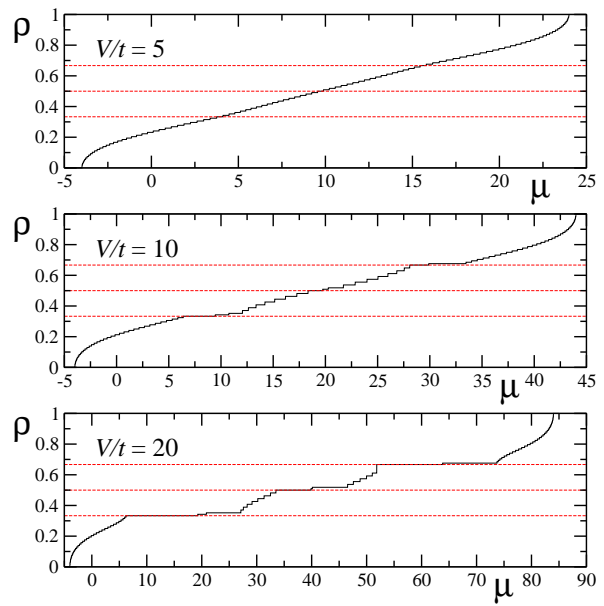}
\caption{\label{fig:rhovsmu}
(Color online) Boson density $\rho$ as a function of the chemical potential 
$\mu$ for three different values of the interaction $V/t$. Calculations have 
been done on a lattice with $L=108$ sites and open boundary conditions.} 
\end{figure}

\begin{figure}
\includegraphics[width=0.90\columnwidth]{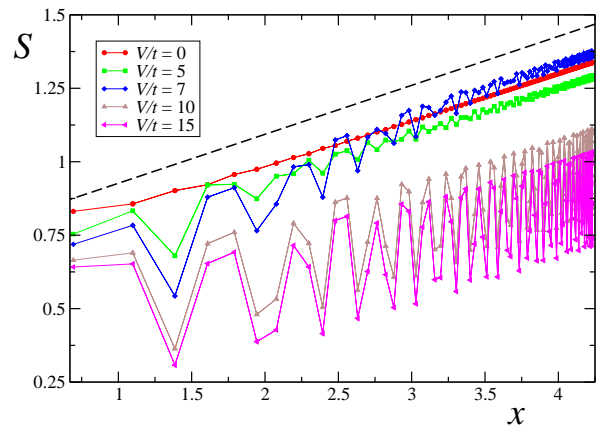}
\caption{\label{fig:central}
(Color online) Entanglement entropy $S$ as a function of the (reduced) block
length $x$, for $\rho=5/12$ and for different values of the interaction 
strength $V/t$. The dashed line indicates the slope $1/6$ (equivalent 
to $c=1$). The number of sites is $L=216$.}
\end{figure}

\begin{figure}
\includegraphics[width=0.90\columnwidth]{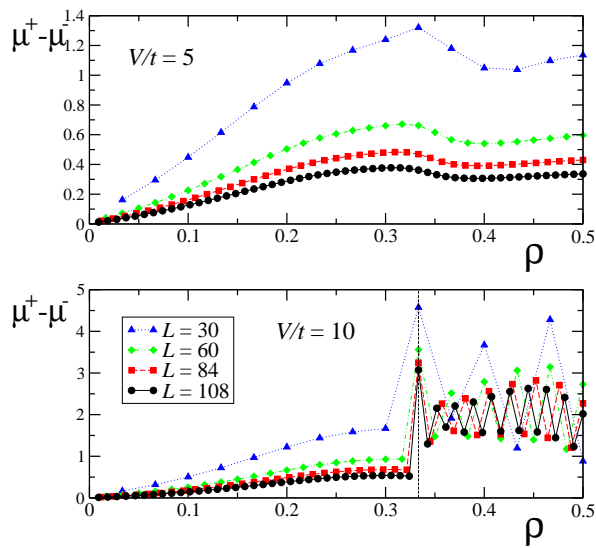}
\caption{\label{fig:mupiumumeno}
(Color online) Difference between the energy needed to add a particle
$\mu^{+}=E(M+1)-E(M)$ and the one needed to remove it $\mu^{-}=E(M)-E(M-1)$ 
(where $E(M)$ is the energy of $M$ particles) as a function of the density 
$\rho$, for different sizes of a system with $V/t=5$ (upper panel) and 
$V/t=10$ (lower panel).} 
\end{figure}

\begin{figure}
\includegraphics[width=0.90\columnwidth]{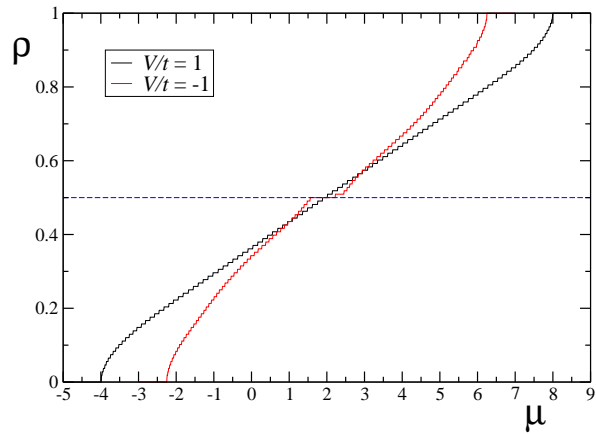}
\caption{\label{fig:pmt}
(Color online) Boson density $\rho$ as a function of the chemical potential
$\mu$ for positive and negative values of the hopping parameters.
Calculations have been done on a system with $L=108$ sites with open boundary 
conditions. The existence of a plateau at $\rho=1/2$ is clear for $t=-1$, 
whereas no plateau is present for $t=1$.} 
\end{figure}

\section{Results}\label{sec:results}

Here, we present our numerical results on the two-leg ladder. The standard 
finite-size DMRG algorithm has been adopted,~\cite{white} fixing the total 
number of sites $L$ and bosons $M$. Systems of up to $L=216$ sites have been 
simulated, keeping $m \sim 300$ states and performing $n_{sw} \sim 6$ sweeps, 
in order to reach convergence both in the energies and in the measure of 
observables. Open-boundary conditions have been used, see 
Fig.~\ref{fig:ladders}(b). The latter choice determines small asymmetries in 
the phase diagram when comparing $\rho$ and $(1-\rho)$, which vanish in the 
thermodynamic limit. In addition, DMRG results have been compared to Lanczos 
diagonalizations up to $L=36$ sites with periodic boundary conditions.

In Fig.~\ref{fig:rhovsmu}, we report the behavior of the boson density $\rho$
as a function of the chemical potential $\mu$: for each $\mu$, the 
corresponding density $\rho=M/L$ is obtained by minimizing the free energy 
$E(M)-\mu M$ with respect to $M$. For small interactions, on any finite size, 
$\rho(\mu)$ displays small steps for every value of $M$; therefore, the ground
state is always gapless, the density being a smooth function of the chemical 
potential $\mu$. For $\rho<1/3$ (and $\rho>2/3$), the system can be described
by a one-component Tomonaga-Luttinger liquid, where the low-energy excitations
are free massless bosons $(\phi,\theta)$ and its central charge is 
$c=1$.~\cite{giamarchi} The latter quantity can be numerically measured through
the entanglement entropy:
\begin{equation}
S(l) = - {\rm Tr}_{\Omega} \left [ \rho_{rdm}(l) \ln \rho_{rdm}(l) \right ],
\end{equation}
where $\rho_{rdm}(l)$ is the reduced density matrix for the subsystem $\Omega$,
containing all the sites from $1$ to $l$, and is defined by:
\begin{equation}
\rho_{rdm}(l) = {\rm Tr}_{\bar \Omega} |\Psi_0\rangle \langle \Psi_0|.
\end{equation}
Here $|\Psi_0\rangle$ is the ground state wave function and ${\bar \Omega}$
defines the environment (all sites from $l+1$ to $L$). The central charge
can be obtained from the logarithmic divergence of $S(l)$:~\cite{wilczek,cardy}
\begin{equation}
S(l) = \frac{c}{6} \ln l + \dots,
\end{equation}
which is valid in the thermodynamic limit and $l \gg 1$. In finite systems,
it is useful to consider:~\cite{hikihara} 
\begin{equation}
x = \ln \left [ \frac{L}{\pi} \sin \left ( \frac{\pi l}{L} \right ) \right ],
\end{equation}
instead of $\ln l$. Since the reduced density matrix $\rho_{rdm}(l)$ is 
directly available by DMRG calculations, the central charge $c$ can be easily 
computed, by fitting the linear slope of $S(l)$ as a function of $x$.
We verified that $c=1$ in the low-density regime $\rho<1/3$ and at any 
interaction strength.

The intermediate boson density at weak coupling requires a deeper discussion.
Indeed, a naive analysis based upon the band structure (and the connection to 
spin-less fermions through the Jordan-Wigner transformation~\cite{jordan}) 
would suggest that the low-energy physics is described by a two-component
Tomonaga-Luttinger liquid, with $c=2$. However, we find that $c=1$ also in this
regime, see Fig.~\ref{fig:central}. The limiting case with $V=0$ corresponds to 
``non-interacting'' hard-core bosons with unfrustrated hopping amplitude,
which are expected to give rise to a standard quasi-condensed quantum liquid.
Therefore, spin-less fermions and hard-core bosons are inherently different 
in this regime of densities (e.g., the Jordan-Wigner string caused by hopping
along the legs plays a relevant role in this zig-zag geometry). We mention that
in the case with frustrated hopping (equivalent to the antiferromagnetic spin 
XY model), the ground state is expected to develop a chiral order that breaks 
a discrete ${\cal Z}_2$ symmetry.~\cite{nersesyan,lecheminant} 
Similarly, a chiral phase has been also predicted for isotropic frustrated 
spin chains in presence of a magnetic field.~\cite{hikihara,vekua2,mcculloch} 
On the contrary, for our unfrustrated model, we do not find any evidence of 
symmetry breaking for $V \ge 0$, as can be seen from Lanczos 
spectra at low-energy (not shown). Therefore, we conclude that there is a 
fundamental difference between frustrated and unfrustrated bosons at these 
densities: for the former case, a gapless state with broken ${\cal Z}_2$ 
symmetry is expected, while, in the latter one, a pure gapless 
Tomonaga-Luttinger liquid is realized.

By increasing the ratio $V/t$, two plateaus emerge at $\rho=1/3$ and $2/3$, 
for $V/t \gtrsim 8$, indicating the stabilization of insulating phases with a 
finite excitation gap, see Fig.~\ref{fig:rhovsmu}. These states can be 
adiabatically connected to the solid phases that have been discussed in 
Sec.~\ref{sec:preliminary}. Remarkably, the stabilization of crystalline phases
at commensurate fillings is accompanied by a significant modification of the 
intermediate phase at $1/3<\rho<2/3$. Indeed, whereas for $V/t \lesssim 8$ the 
liquid phase displays a standard gapless behavior, with a vanishing excitation
energy when adding or removing a single boson, the strong-coupling phase is
instead gapped for single-particle excitations, as clearly shown in 
Fig.~\ref{fig:mupiumumeno}, where the difference between $\mu^{+}=E(M+1)-E(M)$
and $\mu^{-}=E(M)-E(M-1)$ is reported. In the strong-coupling regime, for 
$\rho<1/3$, $\mu^{+}-\mu^{-}$ is a smooth function of $\rho$ that goes to zero
in the thermodynamic limit, while for intermediate densities, this quantity 
remains finite, indicating the presence of a finite gap in the 
single-particle excitations. However, such a strong-coupling phase is not 
insulating, since excitations of pairs of bosons are still gapless. The plot 
of $\rho(\mu)$ shown in Fig.~\ref{fig:rhovsmu} supports this interpretation 
showing that, on any finite-size system, steps twice as big as in the 
standard Tomonaga-Luttinger liquid occur. The emergence of the two-boson bound 
state is simply due to a potential energy gain and does not depend on the sign 
of $t$: as previously noticed, even in the classical limit ($t=0$), the binding 
energy $\Delta=E(M)+E(M+2)-2\,E(M+1)$ is negative in this part of the phase 
diagram (see Sec.~\ref{sec:preliminary}). Not surprisingly, a similar behavior
has been also found in the $J_1{-}J_2$ Heisenberg model in presence of a 
magnetic field.~\cite{vekua,hikihara,okunishi2} In spin systems, the effect, 
which was called ``even-odd'', is characterized by the existence of two-magnon
excitations in the region of weakly coupled antiferromagnetic chains.
This phase may be described by a low-energy theory with two 
bosonic fields $(\phi_n,\theta_n)$, with $n=1,2$, which give rise to 
symmetric and antisymmetric combinations; the presence of relevant 
interactions leads to the opening of an energy gap in the antisymmetric 
channel, implying a two-boson bound state.~\cite{vekua,hikihara}

\begin{figure}
\includegraphics[width=0.90\columnwidth]{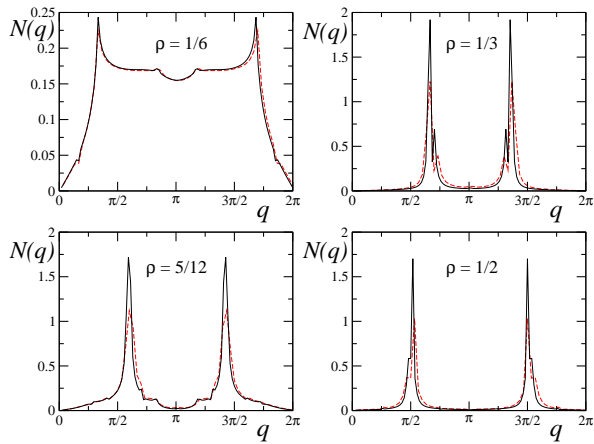}
\caption{\label{fig:sq}
(Color online) Density structure factor $N(q)$ for different boson densities.
Calculations have been done on lattices with $L=108$ (solid curves) and $L=60$
(dashed curves) for $V/t=20$.} 
\end{figure}

\begin{figure}
\includegraphics[width=0.90\columnwidth]{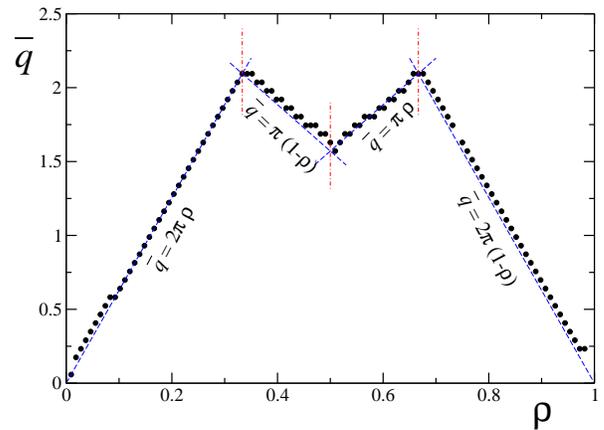}
\caption{\label{fig:qmax}
(Color online) Position of the peak ${\bar q}$ of the density-density
correlations as a function of the density. The value of ${\bar q}$ does not 
depend upon the interaction strength $V/t$.}
\end{figure}

\begin{figure}
\includegraphics[width=0.90\columnwidth]{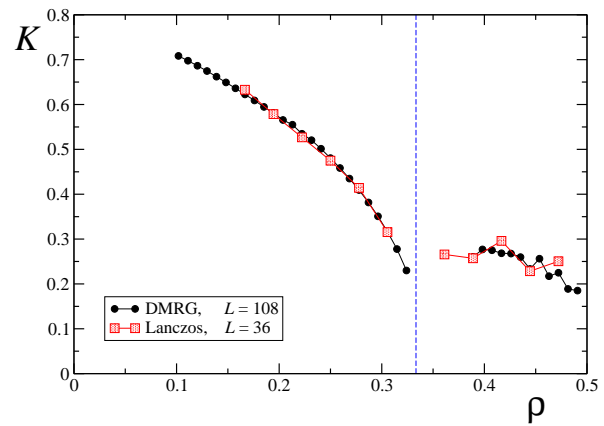}
\caption{\label{fig:Krho}
(Color online) Tomonaga-Luttinger parameter $K$ as a function of the boson
density $\rho$ for $V/t=10$.}
\end{figure}

We would like to mention that the solid at $\rho=1/2$ appears to be less stable 
than the other two at $1/3$ and $2/3$. Indeed, we start to detect its existence
for $V/t=10 \div 11$, although the width of the plateau in a $\rho(\mu)$ 
plot is still fairly small. In Fig.~\ref{fig:rhovsmu}, we show that eventually
a clearly insulating phase exists for $V/t=20$. On the contrary, a very stable
solid phase for $\rho=1/2$ may be obtained by changing sign of the hopping 
parameters; in this case, the model is equivalent to the anisotropic 
$J_1{-}J_2$ Heisenberg chain, which shows a dimerized phase at zero 
magnetization.~\cite{affleck,hikihara2} In Fig.~\ref{fig:pmt}, we report a 
comparison of the curves $\rho(\mu)$ for positive and negative hopping 
parameters and $V/|t|=1$. Only when $t=-1$ a clear plateau is present,
indicating that change in the sign of the hopping parameters has a dramatic 
effect in the stabilization of the insulating phase at half filling. 

\begin{figure}
\includegraphics[width=0.90\columnwidth]{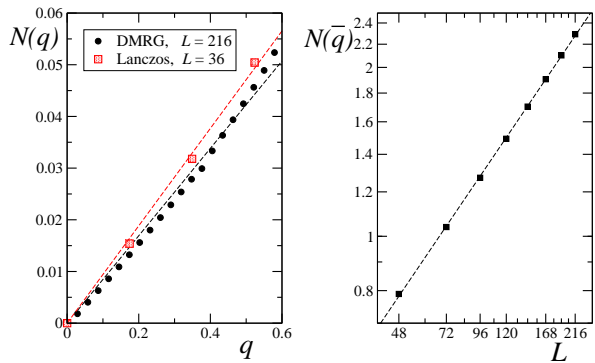}
\caption{\label{fig:sq5_12}
(Color online) DMRG results for the density-density correlation functions on
systems with open boundary conditions. Left panel: small-$q$ behavior of 
$N(q)$ for a $216$-site system. Data for $36$ sites with periodic 
boundary conditions obtained by Lanczos diagonalizations are also shown. 
The boson density is $\rho=5/12$, while $V/t=10$. 
A linear fit of numerical DMRG data gives $K=0.27 \pm 0.01$, while 
for Lanczos data $K=0.29 \pm 0.02$. Apart from the system sizes, 
the small discrepancies could be ascribed to different boundary 
conditions imposed in the two cases. Right panel: size scaling of the peak of 
the structure factor $N({\bar q}) \propto L^\alpha$. The exponent 
$\alpha=0.71 \pm 0.01$ is in good agreement with the value $1-K$ obtained from
the previous fits. Dashed lines are fits of the numerical results.} 
\end{figure}

The nature of the ground state can be better characterized by studying 
the density-density correlation function. In the case with open-boundary
conditions, used within DMRG calculations, we define:
\begin{equation}
N(r,r^\prime)= \langle n_{r} n_{r^\prime} \rangle - 
\langle n_{r} \rangle \langle n_{r^\prime} \rangle,
\end{equation}
which depends separately on $r$ and $r^\prime$. It is convenient to perform 
the Fourier transform
\begin{equation}
N(q)= \frac{2}{L+1} \sum_{r,r^\prime} \sin(q \, r) \sin(q \, r^\prime) 
N(r,r^\prime),
\end{equation}
with $q=2\pi n/(L+1)$ and $n=1,\dots,L$. With periodic-boundary conditions,
implemented with Lanczos algorithm, correlation functions only depend upon
the distance $|r-r^\prime|$, and the standard Fourier transform can be
employed. In Fig.~\ref{fig:sq}, we show the behavior of the density structure 
factor $N(q)$ at $V/t=20$ for different densities. For $\rho<1/3$, the ground 
state is described by a standard one-component Tomonaga-Luttinger liquid, for 
which (in a translationally invariant system):
\begin{equation}\label{eq:TLL1}
\langle n_{r} n_{0} \rangle =  \rho^2 - \frac{K}{2\pi^2r^2} + 
A \frac{\cos(2\pi \rho \, r)}{r^{2K}} + \dots
\end{equation}
where $A$ is a numerical constant and $K$ is the Tomonaga-Luttinger parameter 
that describes the low-energy model.~\cite{giamarchi} $N(q)$ displays a peak 
at ${\bar q}=2\pi \rho$ (see Fig.~\ref{fig:qmax}), in agreement with a 
single-band picture with $k_F=\pi \rho$. No long-range order is implied, since
the finite-size peak grows slower than $L$: $N({\bar q}) \sim L^{1-2K}$. 
The one particle density matrix decays as $1/r^{1/2K}$. Therefore, the dominant
correlations have a charge-density wave (CDW) character for $K<1/2$, while for 
$K>1/2$ a divergence in the $q=0$ occupation number occurs (i.e., 
quasi-condensation). 

A convenient way to compute the Tomonaga-Luttinger parameter $K$ is via the 
small-$q$ behavior of $N(q)$. From Eq.~(\ref{eq:TLL1}) we easily obtain 
$N(q)=\frac{K}{2\pi} \, q$. In Fig.~\ref{fig:Krho}, we report the resulting $K$
as a function of the density $\rho$ for $V/t=10$ (also obtained by Lanczos 
diagonalizations on $L=36$ sites). At low density, bosons are in a 
quasi-condensed state but, close to quarter filling, the crossover to a CDW 
phase takes place. 

At the commensurate density $\rho=1/3$ and sufficiently large $V/t$, 
true long-range order sets in and $N({\bar q}) \sim L$, together with a 
quadratic behavior of $N(q)$ at small momenta, i.e., $N(q) \sim q^2$. 
When approaching $\rho=1/3$ by varying the boson density, the parameter $K$
is expected to remain finite, with a value that does not depend upon the
interaction strength and is twice the limiting value obtained at constant 
density ($\rho=1/3$) when $V/t$ tends to the transition value from 
below.~\cite{giamarchi2}
We mention that, in the classical limit $t=0$ by using the results of
Ref.~\onlinecite{gomez}, we find that $K=1/9$ for $\rho \to 1/3$, which may be
compatible with the strong renormalization that is observed in our numerical
results for $V/t=10$ when approaching the insulating phase, see
Fig.~\ref{fig:Krho}.

For intermediate boson densities, i.e., $1/3<\rho<1/2$, and strong 
interactions the model is described by a two-component boson field 
with a gap in the antisymmetric channel. In this case
\begin{equation}\label{eq:TLL2}
\langle n_{r} n_{0} \rangle =  \rho^2 - \frac{K}{\pi^2r^2} + 
A^\prime \frac{\cos[\pi(1-\rho) \, r]}{r^{K}} + \dots
\end{equation}
where $A^\prime$ is a numerical constant and $K$ is the Tomonaga-Luttinger
parameter in the effective low-energy model of the gapless bosonic 
field. Pair condensation can be investigated trough the correlation function 
\begin{equation}\label{eq:TLL2pair}
\langle b^\dag_{r} b^\dag_{r+1} b_{0} b_{1} \rangle =  
\frac{B}{r^{1/K}} + \dots 
\end{equation}
where $B$ is a numerical constant. In this regime, $N(q)=\frac{K}{\pi} q$ and 
the peak of the structure factor shifts to ${\bar q}=\pi(1-\rho)$, see 
Fig.~\ref{fig:qmax}. Again there is no true long-range order, since 
$N({\bar q})$ diverges as $L^{1-K}$. By fitting the small-$q$ part of the 
structure factor, we are able to extract the behavior of $K$ also in this part
of the phase diagram, see Fig.~\ref{fig:Krho} for the case of $V/t=10$. 
A fairly good agreement is obtained by comparing the power law divergence of 
the peak in $N(q)$ with the expected exponent $1-K$.  In Fig.~\ref{fig:sq5_12},
we report these results for $\rho=5/12$ and $V/t=10$. In this phase, CDW 
correlations dominate and coexist with power-law pairing correlations, which 
however do not lead to (quasi) pair condensation because $K<1$ in the whole
density interval. 

Finally, for $\rho=1/2$ and large $V/t$, the ground state has again crystalline
order with $N({\bar q}) \sim L$ and $N(q) \sim q^2$.

\section{Conclusions}\label{sec:conclusions}

In this paper, we have studied hard-core bosons with unfrustrated hopping and
nearest-neighbor interaction on a frustrated zig-zag ladder. For small $V/t$,
the ground state is a gapless Tomonaga-Luttinger liquid for all densities.
For $\rho<1/3$ (and $\rho>2/3$), the density-density correlations have a 
peak for ${\bar q}=2 \pi \rho$. This fact suggests that a fermionic band
picture (originating from the standard Jordan-Wigner
transformation~\cite{jordan}) can be correct and the Jordan-Wigner string is 
a small perturbation to free spin-less fermions. For intermediate densities 
$1/3<\rho<2/3$, $N(q)$ shows a peak at ${\bar q}=\pi (1-\rho)$, still having
one gapless mode (e.g., $c=1$). This fact contrasts the naive expectation
based upon the Jordan-Wigner transformation. Moreover, we do not find any
evidence of discrete ${\cal Z}_2$ symmetry breaking, as found in the case with
frustrated hopping.~\cite{nersesyan,lecheminant} In this case, for weak
interactions, the ground state is a pure quantum liquid, with a single gapless
mode. By increasing $V/t$, three solid phases appear at commensurate boson 
densities, first at $\rho=1/3$ and $2/3$ (for $V/t \simeq 8$) and then at 
$\rho=1/2$ (for $V/t \simeq 10$): these are insulating phases with long-range 
order in the density profile. For $1/3<\rho<2/3$ there is CDW phase, which has 
gapped single-particle excitations but gapless excitations for pairs of bosons.
This phase, which has been also found in frustrated antiferromagnetic spin
models, can be described by a two-component Tomonaga-Luttinger model with a
finite gap in the antisymmetric sector.

\acknowledgments

We thank M. Fabrizio for many important discussions during the whole duration
of the project. F.B. also thanks T. Giamarchi for interesting suggestions
in Santa Barbara, during the program ``Disentangling Quantum Many-body Systems:
Computational and Conceptual Approaches''. F.B. wants to acknowledge the
fact that this research was supported in part by the National Science 
Foundation under the Grant No. NSF PHY05-51164.
D.R. acknowledges support from EU through the project SOLID, under the
grant agreement No. 248629.
The DMRG code released within the PwP project (www.dmrg.it) has been used.


\begin{thebibliography}{99}

\bibitem{blochrev} I. Bloch, J. Dalibard, and W. Zwerger, \rmp {\bf 80}, 885 
   (2008)
\bibitem{lewenstein} For a recent review of cold atomic gases in optical 
   lattices, see for example, M. Lewenstein, A. Sanpera, V. Ahufinger, 
   B. Damski, A. Sen De, and U. Sen, Adv. Phys. {\bf 56}, 243 (2007).
\bibitem{bloch} M. Greiner, O. Mandel, T. Esslinger, T.W. Hansch, I. Bloch,
   Nature (London) {\bf 415}, 39 (2002).
\bibitem{esslinger} R. Jordens, N. Strohmaier, K. Gunter, H. Moritz,
   T. Esslinger,  Nature (London) {\bf 455}, 204 (2008).
\bibitem{fisher} M.P.A. Fisher, P.B. Weichman, G. Grinstein, and D.S. Fisher, 
   \prb {\bf 40}, 546 (1989).
\bibitem{jaksch} D. Jaksch, C. Bruder, J.I. Cirac, C.W. Gardiner, and 
   P. Zoller, \prl {\bf 81}, 3108 (1998).
\bibitem{demler} L.M. Duan, E. Demler, and M.D. Lukin, {\bf 91}, 090402 (2003).
\bibitem{paredes} B. Paredes, A. Widera, V. Murg, O. Mandel, S. Folling, 
   I. Cirac, G.V. Shlyapnikov, T.W. Hansch, and I. Bloch, Nature (London) 
   {\bf 429}, 277 (2004).
\bibitem{kinoshita} T. Kinoshita, T. Wenger, and D.S. Weiss, Science {\bf 305},
   1125 (2004).
\bibitem{oberthaler} M. Albiez, R. Gati, J. Folling, S. Hunsmann, M. Cristiani,
   and M.K. Oberthaler, \prl {\bf 95}, 010402 (2005).
\bibitem{giamarchi} See for example, T. Giamarchi, {\em Quantum Physics
   in One Dimension} (Oxford University Press, Oxford, 2004).
\bibitem{lanczos} K. Lanczos, J. Res. Natl. Bur. Stand {\bf 45}, 225 (1950).
\bibitem{white} S.R. White, \prl {\bf 69}, 2863 (1992); \prb {\bf 48}, 10345
   (1993).
\bibitem{okunishi} K. Okunishi and T. Tonegawa, J. Phys. Soc. Jpn. {\bf 72},
   479 (2003).
\bibitem{vekua} F. Heidrich-Meisner, A. Honecker, and T. Vekua, \prb {\bf 74},
   020403(R) (2006). 
\bibitem{hikihara} T. Hikihara, T. Momoi, A. Furusaki, and H. Kawamura,
   \prb {\bf 81}, 224433 (2010).
\bibitem{wilczek} C. Holzhey, F. Larsen, and F. Wilczek, Nucl. Phys. B 
   {\bf 424}, 443 (1994).
\bibitem{cardy} P. Calabrese and J. Cardy, J. Stat. Mech.: Theory Exp.
   (2004) P06002.
\bibitem{jordan} P. Jordan and E. Wigner, Z. Phys. {\bf 47}, 631 (1928).
\bibitem{nersesyan} A.A. Nersesyan, A. O. Gogolin, and F.H.L. Essler, \prl
   {\bf 81}, 910 (1999).
\bibitem{lecheminant} P. Lecheminant, T. Jolicoeur, and P. Azaria, \prb 
   {\bf 63}, 174426 (2001)
\bibitem{vekua2} A. Kolezhuk and T. Vekua, \prb {\bf 72}, 094424 (2005).
\bibitem{mcculloch} I.P. McCulloch, R. Kube, M. Kurz, A. Kleine, 
   U. Schollwock, and A.K. Kolezhuk, \prb {\bf 77}, 094404 (2008).
\bibitem{okunishi2} K. Okunishi and T. Tonegawa, \prb {\bf 68}, 224422 (2003).
\bibitem{affleck} S.R. White and I. Affleck, \prb {\bf 54}, 9862 (1996).
\bibitem{hikihara2} T. Hikihara, M. Kaburagi, and H. Kawamura, \prb {\bf 63},
   174430 (2001).
\bibitem{giamarchi2} T. Giamarchi and A.J. Millis, \prb {\bf 46}, 9325 (1992);
   T. Giamarchi, Physica B {\bf 230-232}, 975 (1997).
\bibitem{gomez} G. Gomez-Santos, \prl {\bf 70}, 3780 (1993).
\end{thebibliography}
\end{document}